\documentclass[a4paper]{article}

\def\half{{1\over 2}}

\newcommand{\pdr}{\partial}
\newcommand{\beqs}{\begin{eqnarray}}
\newcommand{\eeqs}{\nonumber\end{eqnarray}}
\newcommand{\beq}{\begin{eqnarray}}
\newcommand{\eeq}{\end{eqnarray}}

\newcommand{\eps}{\epsilon}

\def\m#1{${#1}$}

\def\tr{\;{\rm tr}}

\def\div{\;{\rm div}\;}
\def\curl{\;{\rm curl}\;}

\def\ben{\begin{enumerate}}
\def\een{\end{enumerate}}

\begin{document}
\def\m#1{$#1$}

\centerline{\bf  Fuzzy Fluid Mechanics in Three Dimensions}

\begin{center}
{\sf S. G. Rajeev}\\
Department of Physics and Astronomy\\ 
Department of Mathematics\\
University of Rochester\\
  Rochester NY 14627
\end{center}

\centerline{\bf Abstract}

We  introduce a rotation invariant  short distance cut-off in the theory of an ideal fluid in three space dimensions, by requiring momenta to take values in a  sphere. This leads to an  algebra of functions in position space that  is non-commutative. Nevertheless it is possible to find appropriate analogues of the Euler equations of an ideal fluid. The system still has a hamiltonian structure. It is hoped that this will be useful in the study of possible singularities in the evolution of Euler ( or Navier-Stokes )equations in three dimensions.

\section{Introduction}
A square-integrable  function in Euclidean space can be represented as  a Fourier transform
$$
f(x)=\int \tilde f(k)e^{2\pi ik\cdot x} dk
$$
The wavenumber \m{k} also takes values in Euclidean space, which is naturally thought of as the dual of the original space. If the function is smooth its Fourier transform will decay faster than any power in the dual space. 

If we consider instead  functions on a lattice \m{f:Z^d\to C},  the Fourier transform is a function on the torus $\tilde f:T^d\to C$. In the language of solid state physics, the fundamental domain of this torus is the `Brillouin zone' of momenta of an electron in a periodic potential. Because there is a smallest possible length ( the distance between nearest neighbors in the lattice) there is a largest possible momentum (half the diameter of the Brillouin zone).

Thus there is a reciprocal relation between the smallest distance allowed in space and the largest allowed wavenumber. It is analogous to the uncertainty principle of quantum mechanics. Indeed,  it  {\em is }
the uncertainty principle, once it is accepted that particles are represented by waves.

When studying a partial differential equation it is often useful to impose such a smallest possible length scale, at the cost of introducing some non-locality in the problem. This happens if we replace space by a lattice to discretize the PDE to solve it numerically. As noted above, this imposes a limit on the magnitude of the largest possible wavenumber. In quantum field theory, such a cut-off in momentum (`regularization') is needed to control the divergences that arise inevitably. After this regularization, we would study the dependence  of the parameters ( coupling constants) on the cutoff  implied by the condition that physical observables be independent of the cutoff. This is the still mysterious method of `renormalization'. Applications of quantum field theory to high energy physics as well as critical phenomena require cutoff procedures that preserve the symmetries of the original field theory.

 The lattice method is not always the best regularization, as it breaks rotation invariance. Also, replacing space by a lattice of points introduces a lack of smoothness of functions. Consequently, numerical methods of solving PDEs  suffer from spurious instabilities.

Is it possible to introduce a smallest possible length in space without breaking rotation invariance, and while maintaining smoothness of functions in space? Some proposals of this kind have appeared in theories of quantum gravity and in particle physics\cite{DSR,fuzzyphysics,fuzzylattice} (where the symmetry of interest is Lorentz invariance rather than rotation invariance). The price we pay for this is a kind of fuzzyness in space, where its co-ordinates become non-commutative.  It is no longer possible to locate particles with infinite precision; the algebra of  functions on space is replaced with a non-commutative algebra. These techniques could be  useful  in the theory of PDEs and QFT; and also of practical use in solving PDEs numerically. 

In this paper we will investigate the consequences of introducing such a smallest possible length scale into fluid mechanics in three dimensions. There is some earlier work on two dimensional fluid mechanics\cite{Zeitlin,Abarbanel,FairlieZachos,Dowker,proofregularity}. 
Although there are some situations where it is possible to limit  fluid flow to two dimensions, the vast majority of phenomena of interest are in three dimensions. Some of the deepest problems in physics ( turbulence) and in mathematics (existence of solutions to Navier-Stokes equations) are in three dimensional fluid mechanics. A fundamental phenomenon is that information flows into large structures in  the fluid from small distance scales (causing apparent randomness of the large scale degrees of freedom) while energy flows into small scales (dissipation due to turbulence and viscosity). Mathematically\cite{tao}, three dimensional  Navier-Stokes is a `hard' PDE because it is `supercritical': the nonlinearities become stronger  at small distance scales, making it impossible to know using present techniques whether solutions remain smooth for all time. Thus it is crucial to understand the scale dependence of fluid mechanics. 

Experience from quantum field theory suggests that we must first replace the Navier-Stokes equations with a `regularized' version, in which there is a short distance cutoff. It should be possible to understand the regularity of this cutoff version. As pointed out by Tao,
the problem of establishing  is now shifted to studying the limit as this cutoff goes to zero. The experience of quantum field theory suggests that a cutoff that preserves rotation invariance is needed to have a `renormalization' theory  of the limit as the cutoff  is taken to zero. At least it is a worthwhile avenue of study.

While removing the cutoff  is a great mathematical challenge, the cutoff theory itself could of some interest  in physics. After all, the equations of fluid mechanics are an approximation valid for an average of  a large number  molecules. A `fuzzy' version of fluid mechanics would describe even larger scale motion, which averages over  fluid elements themselves. Such a `mesoscopic' theory may be what we need to understand many physical phenomena, such as the stability of large vortices. Because we averaged out fluid motion, the effective theory of large scale motion can be non-local; i.e., the we get integro-differential rather then differential equations.

Computational Fluid Dynamics is important to many engineering  applications from weather prediction to  the design of aircraft. Typically\cite{Patankar}, space is divided into a finite number of cells.  The PDEs are turned into finite difference equations that are solved numerically. If the size of the cell can be made small enough this can give a good approximation to the real  flow. However, the number of cells is limited by the memory of the computer. If the region of space is large  the size of a cell can be too large. For example in weather prediction, a cell is several kilometers in size. This means not only that you miss phenomena within such cells, but also that predictions are limited in time. Given enough time the small scale will affect the large scale motion. In the case of the atmosphere the limit is about ten days beyond which predictions of the weather become unreliable with even the largest computers.

Thus, a method that imposes a smallest possible length, and a largest possible wavenumber,  without breaking symmetries could help us  in mathematical, physical and engineering approaches to fluid mechanics.

Numerical simulations of quantum field theory\cite{fuzzylattice}  by the non-commutative regularization sometimes runs into trouble, because of a kind of anomaly. A remnant of non-commutativity survives even in the limit as the regularizaton is removed. Thus, the limiting theory is not what one would have hoped to have originally. It can be asked 
\cite{sgrajeevcom} whether such problems can arise in classical field theories such as fluid mechanics. At least in two dimensions, the answer is no. It has been rigorously established\cite{proofregularity}  that the limit of the solution of the regularized theory ( on a torus) is indeed the solution of Euler's equation. For a  more detailed and updated discussion   of these issues we refer the reader to the author's webpage \cite{sgrajeevcom}.

\section{The Euler Equation in Three Dimensions}

We now summarize the relevant facts about   fluid flow in three dimensional fluid flow \cite{Lamb,rajeevmrst}.

The  Euler equations of an incompressible inviscid fluid
are 
\beq
{\pdr\over \pdr t}v+(v\cdot\nabla)v=-\nabla p, \div v=0.
\eeq
We can eliminate pressure \m{p} by taking a curl. Defining the vorticity
\m{\omega=\curl v} and   
using \m{(v\cdot \nabla)v=\omega\times v+\half \nabla v^2}, we get, 
\beq
{\pdr\over \pdr t}\omega+\curl[\omega\times v]=0.
\eeq
We can regard \m{\omega} as the basic dynamic variable\cite{ArnoldKhesin} of the system, since
\m{v} determined by it uniquely \footnote{ We assume that the  velocity vanishes at infinity; this is necessary to have finite energy. Or, we could study fluid flow in some domain in $R^3$ with appropriate boundary conditions.} as the solution of the equations
\beq
\div v=0,\quad \curl v=\omega.
\eeq

 \subsection{Clebsch Variables}
 
  It was noticed by Clebsch that any solution to \m{\div \omega=0} can be written as 
 \beq
 \omega=\nabla\lambda\times \nabla \mu
 \eeq
 for a pair of functions \m{\lambda,\mu: R^3\to R}. The  Euler equations take a simple form  in these variables\cite{Lamb}. They are constant along stream lines:
\beq
{\pdr \lambda\over \pdr t}+v\cdot\nabla\lambda=0,\quad
{\pdr \mu\over \pdr t}+v\cdot\nabla\mu=0.
\eeq
Here we are to regard velocity as a function of the Clebsch variables:
\beq
v=P\left(\lambda\nabla\mu\right).
\eeq
Also, \m{P} is the projection  of a vector to its  divergence free part:
\beq
P(u)=u+\nabla\phi,\quad \nabla\cdot u+\nabla^2\phi=0.
\eeq
This follows from the identity
\beq
\nabla\lambda\times\nabla\mu=\nabla\times[\lambda\nabla\mu]=\nabla\times v.
\eeq

 \subsection{Some Technical Remarks}
 \begin{enumerate} 
   \item We will lose some global information about the flow  in this parametrization; for example the `kinetic helicity'  (a  conserved quantity) will vanish if $\lambda,\mu$ are globally defined: \m{\int \omega\times v dx=0.} To avoid this,  we must let \m{(\lambda,\mu)} be co-ordinates on a two dimensional manifold  more general than \m{R^2}. But we postpone such issues for now.

 \item There is a `gauge invariance' to this parametrization: under canonical transformations in the pair of co-ordinates \m{(\lambda,\mu)}, vorticity is unchanged.

 \item J. Marsden and A. Wienstein\cite{MarsdenWeinstein} have given a nice geometrical interpretation of these variables. The space of pairs of functions \m{(\lambda,\mu)} is a symplectic vector space using the contraction \m{<\lambda,\mu>=\int \lambda\mu d^3x} given by the volume form. The group of incompressible diffeomorphisms act on this symplectically. The Clebsch bracket \m{\omega=[[\lambda,\mu]]:=\nabla\lambda\times \nabla \mu} is the moment map of this action. This is anlogous to the formula \m{L=r\times p} for angular momentum.

\item We propose another interpretation that suggests a natural deformation. Let \m{\lambda,\mu:\underline{G}^*\to R} be  pair of functions on the dual of a Lie algebra \m{\underline{G}}. Then \m{d\lambda} and \m{d\mu} can be thought of as functions \m{:\underline{G}^*\to \underline{G}}. Thus it makes sense to define the commutator \m{[d\lambda,d\mu]}. This is a map \m{V\otimes V\to V\otimes\underline{G}}, where \m{V} is the space of real functions on \m{\underline{G}}. As  a vector space we can identify \m{V\equiv U(\underline{G})}, the universal envelope of \m{\underline{G}}. In the present case $G=SU(2)$ and its Lie algebra dual is $R^3$.  In a later publication we will discuss the deformation of this universal envelope into a non-co-commutative Hopf algebra ( "quantum group") which provides a natural regularization of both short-distance and large distance effects.

\end{enumerate}

\subsection{ The Hamiltonian in the Clebsch Parametrization}

Another advantage of the Clebsch variables is that ideal fluid flow can be thought of a hamiltonian system in which they are the canonical co-ordinates:
  \beq
 \{\lambda(x),\mu(y)\}=\delta(x,y),\quad \{\lambda(x),\lambda(y)\}=0=\{\mu(x),\mu(y)\}.
 \eeq
The hamiltonian is just the total kinetic energy of the fluid:
\beq
H=\half<v,v>={1\over 2}<\lambda\nabla\mu|P|\lambda\nabla\mu>.
\eeq
It is not difficult to verify that the Euler equations in Clebsch form follow from this hamiltonian and Poisson bracket.

It will be useful to write the hamiltonian in terms of Fourier transforms,
\beq
\lambda(x)=\int \tilde\lambda(k)e^{2\pi i k\cdot x} dk,\quad \tilde\lambda^*(k)=\tilde\lambda(-k).
\eeq
We get 
\beq
\{\tilde\lambda^*(k),\tilde\mu(p)\}=\delta(k,p)
\eeq
all other Poisson brackets being zero. Also,
\beqs
H&=&\half\int {dk}{dp}{dk'}{dp'} \delta(k+p-k'-p')\cr
& &
\tilde\lambda^*(k')\tilde\mu^*(p')\tilde\lambda(k)\tilde\mu(p)
(2\pi)^2\left[p'\cdot p-{p'\cdot (k+p)(k+p)\cdot p\over |k+p|^2}
\right]
\eeqs

\section{Bounded Momentum Space}

A way of getting a configuration space of finite extend for a particle is to imagine that it is moving on a 3-sphere. This preserves rotation invariance, unlike periodic boundary conditions that put the system in a torus.  In the limit of infinite radius we will get back Euclidean space. Its phase space will be $T^*S^3$; using the group identification $S^3\approx SU(2)$ , we get $T^*S^3\approx S^3\times R^3$.

How will be put a cut-off on largest possible wavenumber? Does it make sense that {\em momentum space} is $S^3$? 

Any function \m{f:R^3\to C} has the Fourier representation
\beq
f(x)=\int \tilde f(k)e^{2\pi ik\cdot x}{dk}.
\eeq
The product of two functions corresponds to the convolution defined through the addition of momenta:
\beq
f_1f_2(x)=\int \tilde f_1(k_1)\tilde f_2(k_2) e^{2\pi i[k_1+k_2]\cdot x}dk_1dk_2.
\eeq

If we change the composition law for momenta the rule for multiplication of functions will change as well. If momenta are valued in the non-abelian group $SU(2)$, the multiplication of functions in space becomes non-commutative.

Suppose we are given a smooth,locally invertible, function  $k:SU(2)\to R^3$ such that $k(1)=0$. We can then get  a cutoff version of the Fourier transform 
\beq
f(x)=\int \tilde f(g)e^{2\pi ik(g)\cdot x} dg
\eeq
Here $dg$ is the  invariant (Haar) measure on $SU(2)$ normalized so that 
\beq
\int dg=1.
\eeq
The inverse of this is
\beq
\tilde f(g)=J(g)\int f(x)e^{-ik(g)\cdot x}dx
\eeq
where \m{J(g)=\det {\pdr k\over \pdr g}} is the Jacobian of the transformation \m{k\to g}.

We also define a $*$-product on functions on $R^3$, induced by the group multiplication on $SU(2)$:
\beq
e^{2\pi ik(g_1)\cdot x}*e^{2\pi ik(g_2)\cdot x}=e^{2\pi ik(g_1g_2)\cdot x}.
\eeq
More explicitly,
\beq
f_1*f_2(x)=\int f_1(y)f_2(z)K(x,y,z)dydz
\eeq
where
\beq
K(x,y,z)=\int dg_1dg_2 e^{2\pi i[k(g_1g_2)\cdot x-k(g_1)\cdot y-k(g_2)\cdot z
]}
\eeq
Associativity of this  multiplication follows from that of group multiplication on $SU(2)$. But it is not commutative.

The momentum operators are, in this picture, just multiplication:
\beq
\hat P_i\tilde f(k)=k_i\tilde f(k).
\eeq
Hence 
\beq
[P_i,P_j]=0.
\eeq
Ordinarily the position operators would be the differentiation with respect to \m{k}. Instead, they  will be the generators of the group action:
\beq
X^i=i\xi^i_j(k){\pdr \over \pdr k_j}
\eeq
so that 
\beq
[P_i,X^j]=i\xi^j_i(P)
\eeq
The components of this vector field on the group are determined in terms of the Maurer-Cartan forms:
\beq
\xi^i_j(k)\omega^j_l(k)=\delta^i_l,\quad g^{-1}dg=\omega_i^jdk_j{i\sigma_i\over 2}.
\eeq
It follows that \m{\xi^i_j(k)=\delta^i_k+{\rm O}(k)}. Thus for momenta small compared to the cutoff, the position and momentum  operators satisfy the usual canonical commutation relations.

We can see that 
The position operators no longer commute; instead they satisfy the commutation relations of the Lie algebra.
\beq
[X^i,X^j]=i\eps^{ij}_kX^k.
\eeq
Therefore it will not be possible to locate the position of a particle with infinite precision. This fuzziness in position means that in effect we are averaging physical quantities over a small region whose size $a$ is a kind of cutoff in space, of the order of the inverse of the diameter of the momentum space.

An example of  such a function $k$ is (the inverse of )  the stereographic co-ordinate system
\beq
g={1-a^2k^2\over 1+a^2k^2}+{2ia\sigma\cdot k\over 1+a^2k^2}.
\eeq
where \m{a} is a constant with the dimensions of length. In these co-ordinates, the point at infinity in the variable $k$ corresponds to $g=-1$. Thus we must require all functions to tend to a constant as $|k|\to \infty$. The formula for the modified
Fourier transform can be written as 

\beq
f(x)=\int \tilde f(k)e^{2\pi ik\cdot x} {a^3dk\over (1+a^2k^2)^3}
\eeq
since  $dg={a^3dk\over (1+a^2k^2)^3}$ in these co-ordinates.

 If we set 
\beq
\rho(g)=\arccos\left[\half\tr g
\right]
\eeq
we can solve for $k$ to get 
\beq
|k|={1\over a}\tan\half\rho(g),\quad k=-{i\over 2a}{\tan\half\rho(g)\over \sin\rho(g)}\tr g\sigma
\eeq
In the limit as $a\to 0$ the cutoff Fourier transform and the $*$-product reduces to the usual ones. To be specific we will assume from now on that we are using the stereographic parametrization. Up to second order terms in  \m{a},
\beq
g\approx 1+2ai\sigma\cdot k,\quad k(g)\approx -{i\over 4a}\tr g\sigma,\quad k(g_1g_2)\approx k_1+k_2+ak_1\times k_2+
\cdots
\eeq

Other choices of the function $k$ will give fluid equations  that differ by terms higher  order in $a$. Such an  ambiguity  in the choice of regularization is common.

\section{Euler's Equation With Short Distance Cutoff}

It is no problem generalizing the Poisson brackets; they just involve delta functions. 
\beq
\{\tilde\lambda^*(g_1),\tilde\mu(g_2)\}=\delta(g_1^{-1}g_2)
\eeq
The other brackets \m{\{\lambda(g_1),\lambda(g_2)\}} and \m{\mu(g_1),\mu(g_2)\}} remain  zero.

The hamiltonian needs some work.
We should replace \m{k,p,k',p'} by elements of $SU(2)$. It is useful to denote
\beq
<g_1,g_2>={1\over 4a^2}\left[\tr g_1^\dag g_2-2\right]
\eeq
As \m{a\to 0}, $<g_1,g_2>\approx k_1\cdot k_2$.

This way we can get a formula 
\beqs
H&=&\half\int dg_1dg_2dg_1'dg_2'\delta(g_1'g_2'g_1^{-1}g_2^{-1})\cr
& &
\tilde\lambda^*(g_1')\tilde\mu^*(g_2')\tilde\lambda(g_1)\tilde\mu(g_2)
\left[<g_2',g_2>-{<g_2',g_1g_2><g_1g_2,g_2>\over <g_1g_2,g_1g_2>}
\right]
\eeqs

Thus, in principle we have a cutoff version of three dimensional fluid mechanics. Hamilton's equations follow:
\beqs
{\pdr \tilde\mu(g)\over \pdr t}&=&\int dg_2dg_1'dg_2'\delta(g_1'g_2'g_1^{-1}g_2^{-1})\cr
& &
\tilde\lambda^*(g_1')\tilde\mu^*(g_2')\tilde\mu(g_2)
\left[<g_2',g_2>-{<g_2',gg_2><gg_2,g_2>\over <gg_2,gg_2>}
\right]
\eeqs
\beqs
{\pdr \tilde\lambda(g)\over \pdr t}&=&-\int dg_1dg_2dg_1'dg_2'\delta(g_1'g_2'g_1^{-1}g_2^{-1})\cr
& &
\tilde\lambda^*(g_1')\tilde\mu^*(g_2')\tilde\lambda(g_1)
\left[<g_2',g>-{<g_2',g_1g><g_1g,g>\over <g_1g,g_1g>}
\right]
\eeqs

It would be interesting to see if these equations have a smooth time evolution given smooth initial data. 

This is not quite ready for a numerical applications, because we still have an infinite number of variables. We have a cutoff in momentum space but position space is still infinitely large. A cutoff in both spaces is necessary to get a finite number of degrees of freedom. This can be done, but the resulting algebra of functions is not any more related to a group. It is still associative and is related to a quantum group ( Hopf Algebra). We hope to return to these points.

\section{Acknowledgement}
This work was supported in part by the Department of Energy   under the  contract number 
 DE-FG02-91ER40685. I also acknowledge discussions with H. Abarbanel,  A. Amirdjanova, M. Gordina, A. Jordan, J. Hilgert, D. Mumford,T. Wurzbacher and C. Zachos. I also thank M. Panero for a 
 discussion of possible non-commutative anomalies and for drawing my attention to references \cite{fuzzylattice}.

\end{document}